\documentclass[
 aip,
 amsmath,amssymb,
 reprint,%
 nofootinbib
]{revtex4-1}

\usepackage{graphicx}
\usepackage{dcolumn}
\usepackage{bm}
\usepackage{enumerate}

\usepackage{setspace,color,url}
\allowdisplaybreaks[1]

\usepackage[dvipsnames]{xcolor}

\usepackage{graphicx}
\usepackage{soul}
\usepackage{mathrsfs}

\begin{document}



\title{Electrokinetics with mobile surface charges and application to blue energy harvesting}
\title{Interfacial transport with mobile surface charges: conductance, electro- and diffusio- osmotic transport}
\title{Interfacial transport with mobile surface charges and consequences for ionic transport in carbon nanotubes}

\author{Timoth\'ee Mouterde, Lyd\'eric Bocquet}
\affiliation{Laboratoire de Physique Statistique, UMR CNRS 8550, Ecole Normale Sup\'erieure, PSL Research University, 24 rue Lhomond, 75005 Paris, France}
\email{lyderic.bocquet@lps.ens.fr}
\date{26 November 2018}
\begin{abstract}
 In this paper, we explore the effect of a finite surface charge mobility on the interfacial transport: conductance, streaming currents, electro- and diffusio-osmotic flows. We first show  that the surface charge mobility modifies the hydrodynamic boundary condition for the fluid, which introduces a supplementary term depending on the applied electric-field.  In particular, the resulting slip length is found to decrease inversely with the surface charge. We then derive expressions for the various transport mobilities, highlighting that the surface charge mobility merely moderates the amplification effect of interfacial slippage, to the noticeable exception of diffusio-osmosis and surface conductance. Our calculations, obtained within Poisson-Boltzmann framework, highlight the importance of non-linear electrostatic contributions to predict the  small concentration/large charge limiting regimes for the transport mobilities.
 We discuss these predictions in the context of recent electrokinetic experiments with carbon nanotubes. 
%
\end{abstract} 
\maketitle

\section{Introduction}

Electro-osmosis, streaming currents or diffusio-osmosis are canonical examples of electrokinetic transport in soft matter \cite{Hunter}. 
These transport phenomena are all interfacially driven and take their roots within the very first nanometres close to the
channel surfaces  \cite{Bocquet2010}. This yields a high sensitivity of electrokinetic transport to surface details: this both raises considerable difficulties in the description of these intrinsically nanofluidic phenomena, but also brings a considerable
richness, leading in some cases to disruptive innovations {\it e.g.} in energy harvesting \cite{Siria2017}.
Numerous interfacial properties do affect the electrokinetic transport, among which one may quote interfacial hydrodynamic slippage \cite{Joly2004,Joly2006,Balme2015,Huang2016}, ion specificity
\cite{Huang2008}, dielectric anomalies \cite{Bonthuis2012,Bonthuis2016}, charge regulation \cite{Secchi2016PRL,Biesheuvel2016, Uematsu2018}, Stern layer mobility \cite{vanroij2018}, etc.

In this paper we consider the effect of mobile surface charges on the electrokinetic transport, treating the electric double layer at the non-linear Poisson-Boltzmann level.
Usually, surface charge is considered to result from the chemical reactivity at the surface when dipped into water, involving typically acid-base surface
reaction as $ [A-H]_s \rightarrow [A^-]_s + H^+$, where the negative group $[A^-]_s$ remains fixed on the
surface, while proton diffuses in the solution. The surface charge generates an electric double layer at the root of 
the interfacially driven transport phenomena.

Here we consider a situation where the (say) negative groups generating the surface charge do physisorb on the surface, but without covalent bonding with the surface. This is for example the situation of charged groups confined within fluid membranes \cite{Fleck2002,Andelmann2018}, or the surfactant adsorption on liquid-vapor interface, potentially leading to electro-osmotic mobility \cite{Joly2014}.
Here we rather consider as a prototypical example the weak physisorption of (say, negative) ions on a solid surface, so that these keep a lateral mobility while adsorbed on the surface.
The question of ion transport within the so-called 'stagnant layer' in the electric double layer was raised in a number of early works 
\cite{Lyklema1994}, notably in an attempt to explain discrepancies between the surface charge and its electrokinetic counterpart \cite{Lyklema1994,Bonthuis2016}. {However the possibility of more advanced modifications of the hydrodynamic boundary condition as investigated here, and its consequences, was not considered}.
In this context, new insights are provided by experiments on fluid and ionic transport in single carbon nanotubes. These systems were shown to exhibit large hydrodynamic slippage at their surfaces \cite{Secchi2016Nature}, while electrokinetic measurements suggested the presence of a surface charge taking its root in the (weak) adsorption of hydroxide ions on the carbon interface\cite{Secchi2016PRL}. This picture was confirmed recently by {\it ab initio} simulations \cite{Grosjean}, showing furthermore that the OH$^-$ mobility remains large even while physisorbed on the surface, due to low friction on carbon surfaces. 
Similar conclusions were drawn from the ion mobility in molecular confinement  achieved with graphite walls \cite{Mouterde2018}.
These investigations represent therefore a unique opportunity to disentangle the effects of mobility of physisorbed ion and slippage in electrokinetic transport, shedding a new light on the underlying mechanisms at stake. A full understanding requires that consistent formulas for the transport coefficients be obtained
 taking into account all various dynamical mechanisms at play.

In this paper we thus explore how the mobility of the ions constituting the surface charge do modify the electrokinetic transport phenomena. In particular, we show how the finite ion mobility at the surface modifies the hydrodynamic slippage boundary condition at the surface, as well as its effect on electrokinetic transport. We explore specifically conductance, electro-osmosis, streaming current, diffusio-osmotic flow and diffusio-osmotic current.

\section{Modified hydrodynamic boundary condition}

We first show in this section that the surface charge mobility leads to a modification of the hydrodynamic boundary condition at the surface. 

\begin{figure}
    \centering
    \includegraphics[width=3in]{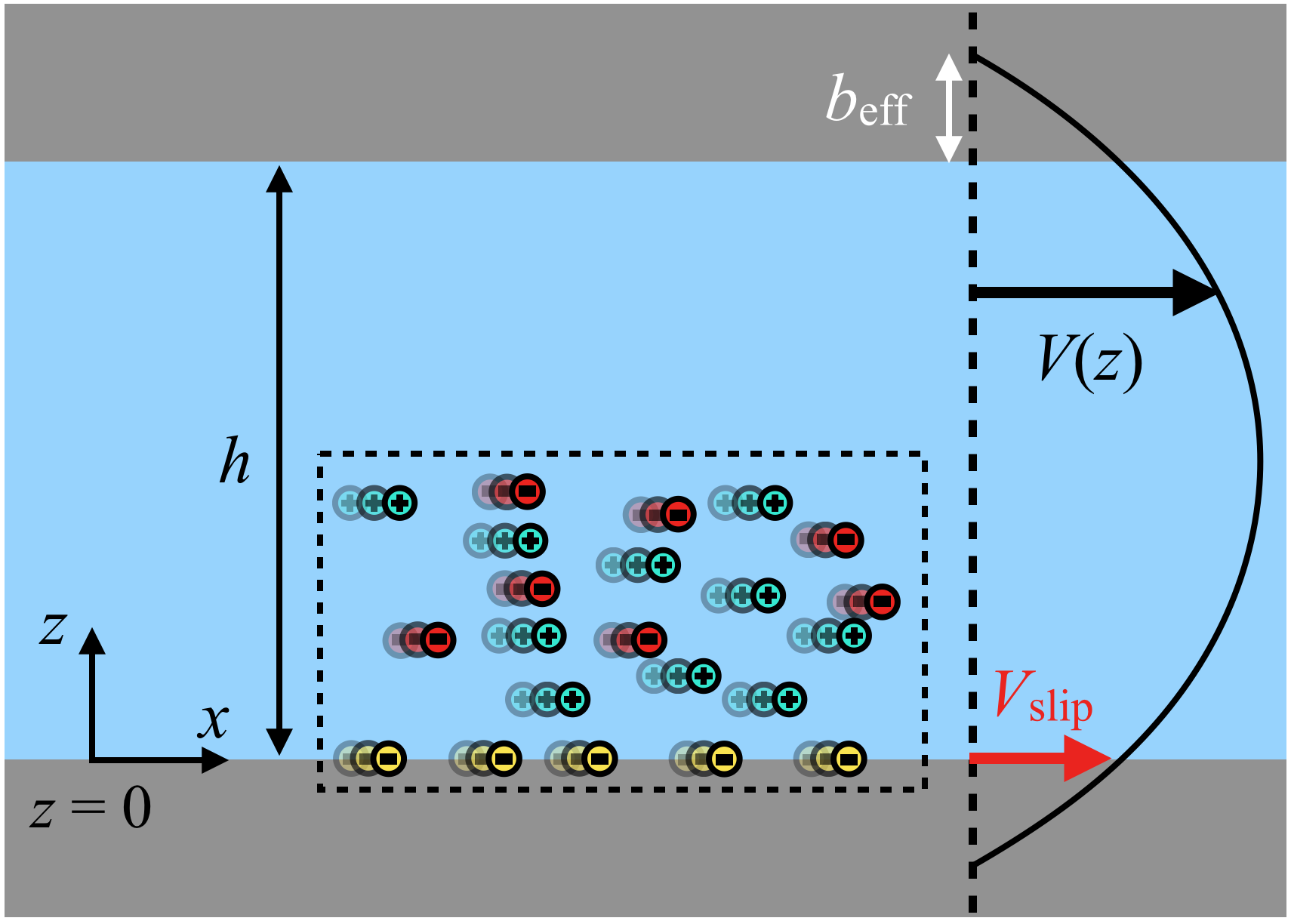}
    \caption{\textsf{Sketch of the considered geometry. In a channel of height $h$, water flows with a velocity $V(z)$ along the $x$-axis, the surface velocity is $V_{\rm slip}$ prescribed by the effective slip length $b_{\rm eff}$. The mobile adsorbed ions are sketched in yellow, while the bulk ions are in blue and red for positive and negative charges, respectively.}}
    \label{fig1}
\end{figure}

We consider a semi-infinite fluid volume and we define $x$ and $z$ as the axis along and perpendicular to the plane, respectively (see Figure \ref{fig1}). We choose $z=0$ at the position of the adsorbed ions. 
We denote the surface charge as $\Sigma$ (C.m$^{-2}$) and assume here for simplicity that it is negative.
This charge is assumed here to result from the weak physisorption of a specific ion species at the surface -- {\it e.g.} OH$^-$ for carbon nanotubes -- , {we treat the possible bulk and adsorbed ions of this specie as two distinct populations with prescribed densities, a possible link between adsorbed and bulk densities will be discussed below.} We assume accordingly a finite lateral mobility of these adsorbed ions.
We also introduce a surface density of the mobile charge as $\Sigma_n=-\Sigma/e$, and the fluid velocity (along the $x$-axis) as $V(z)$.
We consider a monovalent salt, such as potassium chloride KCl with a bulk concentration $C_\infty$ and we assume, except when specifically stated, that in the bulk the two ions have equal diffusivities $D_{+}= D_{^-}$.

We also assume that in the absence of ions, the fluid obeys a partial slip boundary condition at the surface, with a slip
length $b_0$.

\begin{equation}
b_0\times \partial_z V\biggl\vert_{wall} = V_{slip}
\end{equation}
where $V_{slip}=V(z=0)$ is the slip velocity of the fluid at the surface.
Let us now write the force balance on a single physisorbed ion. In the presence of a water flow -
with slip velocity $V_{slip}$ in the adsorbed layer - and in the presence of a tangential electric field $-\partial_x\psi$ (if any),
the ion will experience friction forces due to its relative motion versus both water molecules and wall surface, as well as a direct electric force.
For a (negative) physisorbed ion with lateral velocity $v_-$,
the force balance in the stationary state can thus be written as: 
\begin{equation}
0=-\lambda_s (v_- - V_{slip}) - \lambda_{w} v_- + (-e) (-\partial_x \psi) 
\end{equation}
where $\lambda_w$ and $\lambda_s$ are the friction coefficients of the ion with  wall and  water, respectively.
This leads to a velocity of a physisorbed ion in the presence of a water flow and electric field, in the form:
\begin{equation}
v_-={\lambda_s\over \lambda_s+\lambda_w} V_{slip} + {(-e) \over {\lambda_s+\lambda_w}} (-\partial_x \psi) 
\label{vion}
\end{equation}
The first term accounts for the additional wall friction acting on the adsorbed ions, so that their
velocity may differ from the fluid velocity at the surface, $V_{slip}$.
Note that as expected, the ion velocity vanishes for infinite ion-wall friction $\lambda_w\rightarrow \infty$.

In order to obtain the modified slip boundary condition in the presence of the mobile ions, we now turn to the force balance on the interfacial fluid layer, consisting in the water and physisorbed ions. This force balance takes the form:
\begin{equation}
0=\Sigma (-\partial_x\psi) -  \lambda_0 V_{slip} - \lambda_w \Sigma_n v_- + \eta \partial_z V\biggl\vert_{wall}
\end{equation}
where $\eta$ is the fluid viscosity and $\lambda_0=\eta/b_0$ is the water-wall friction coefficient.
This equation generalizes the standard force balance on a fluid, with the additional friction effects due 
to the physisorbed ions, as well as the effect of a tangential electric field on this charged layer.

Using the previous expression for the ion velocity $v_-$, one gets the 
modified hydrodynamic boundary condition for the fluid in the presence of the physisorbed ions at the interface, undergoing a tangential electric field:
\begin{equation}
 b_{\rm eff} \times \partial_z V\biggl\vert_{wall} = V_{slip}  - \alpha_s{b_{\rm eff}\over \eta} \times {\Sigma}\,(-\partial_x\psi)\biggl\vert_{wall} 
\label{slip}
\end{equation}

where  $\alpha_s={\lambda_s\over \lambda_s+\lambda_w}$ and the effective slip length is
\begin{equation}
b_{\rm eff}= {b_0 \over 1 + \beta_s \times \Sigma_n}
\label{bslip}
\end{equation}
with $\beta_s={1 \over \lambda_0}{\lambda_s\lambda_w\over \lambda_s+\lambda_w}$.
The effective slip length takes into account the ion contribution to the wall friction. Due to the mobility of the ions at the interface, the hydrodynamic boundary condition also involves a new electric term. Note that in this boundary condition, we omitted for simplicity surface charge gradients, which may lead to Marangoni like contributions in the force balance. In the following we will neglect such contributions which we leave for future studies, and rather focus on the effect of finite surface charge mobility.


\section{Conductance, streaming and electro-osmosis}

\subsection{Electrostatics within the electric double layer:  basic reminders}

Before exploring the transport phenomena, we recall several classical resuts of the electric double layer which will be useful for the following calculations, see Ref. \cite{Andelman} for a detailed review.
In the framework of the Poisson-Boltzmann equation, the equilibrium concentration profile of ions takes the Boltzmann form,
$C_{\pm}=C_\infty \exp(\mp \Phi)$, 
with $\Phi= e\psi/k_BT$ the dimensionless electrostatic potential. The potential $\psi$ verifies Poisson's equation:
{
\begin{equation}
\nabla^2\psi = - { e (C_+-C_-)\over \epsilon} - {\Sigma \over \epsilon} \delta (z)
\end{equation}
}
with $\epsilon$ the dielectric permittivity, $e$ the elementary charge and $\delta(z)$ the Dirac distribution.
The resulting Poisson-Boltzmann (PB) equation for the electrostatic potential can be written as:

{
\begin{equation}
\nabla^2 \Phi = \kappa^2 \sinh\Phi + {2 \delta (z) \over \ell_{GC}}
\label{PB}
\end{equation}
}
with $\kappa^2=8\pi \ell_B C_\infty$ the screening factor and  $\ell_B$  the Bjerrum length $\ell_B= e^2/4\pi \epsilon k_BT$. This also introduces 
the Gouy-Chapman length defined as $\ell_{GC}=1/2\pi \Sigma_n \ell_B$, see  \cite{Andelman}.
In the semi-infinite 1D geometry considered here, the PB equation can be integrated to yield:
\begin{equation}
\tanh {\Phi\over 4} = \gamma \exp(-\kappa z)
\end{equation}
with $\gamma = \tanh ({e\psi_0\over 4 k_BT})$ and $\psi_0$ the surface potential. The parameter $\gamma$ obeys the equation $\gamma^2 + 2 {\ell_{GC}\over \lambda_D} \gamma -1 =0$.

Gauss theorem shows that the relation between the (dimensionless) surface potential $\Phi_0=e\psi_0/k_BT$ and the physisorbed surface charge $\Sigma$ 
is $\Sigma = {e\over 2 \pi\ell_B \lambda_D} \sinh {\Phi_0\over 2}$, which can be conveniently rewritten as:
\begin{equation}
\sinh {\vert \Phi_0 \vert\over 2}={1\over \kappa\ell_{GC}}
\label{Surfpot}
\end{equation}

In the following a key parameter will be $\chi=1/\kappa\ell_{GC}$, the ratio between the Debye length $\lambda_D=\kappa^{-1}$ and
the Gouy-Chapman length. This parameter $\chi$ is proportional to the surface charge $\chi \propto \vert\Sigma\vert$.

\subsection{Conductance}

Let us first consider the effect of surface charge mobility on the conductance across a channel.
To simplify the discussion, we consider a rectangular geometry with height $h$, width $w\gg h$, 
and length $L\gg h$. The thickness of the Debye layer is assumed to be small compared to the channel height $ h \gg \lambda_D$.

The current under an applied (tangential) electric field $E_0$ is:
\begin{eqnarray}
&I_{e} = &w\times \int_0^h e(C_+-C_-)(z)\,V(z)dz \nonumber \\
&&+ w\times \int_0^h e^2(\mu_+C_++\mu_-C_-)(z)\,E_0dz +  2w\times \Sigma v_-\nonumber \\
\end{eqnarray}
where the ion mobilities are defined as $\mu_\pm={D_\pm\over k_BT}$, here  assumed to be equal for simplicity 
$\mu=\mu_+=\mu_-$ (which is {\it e.g.} the case of KCl).

We write the expression for the current as the sum of two terms, $I_e= I_{no-slip}+I_{slip}$,
where:
\begin{eqnarray}
&I_{\rm no-slip} = &w\times \int_0^h e(C_+-C_-)(z)\times (V(z) - V_{slip})dz \nonumber \\
&&+ w\times \int_0^h e^2(\mu_+C_++\mu_-C_-)(z)\,E_0\,dz
\end{eqnarray}
accounts for the 'no-slip' contribution to the current and
the 'slip' contribution to the conductance is defined as 
\begin{eqnarray}
&I_{\rm slip} = & w \int_0^h e(C_+-C_-)(z)\,V_{slip}dz+  2w\times \Sigma v_-
\end{eqnarray}

The derivation of the 'no-slip' contribution to the current follows as usual, and we leave details to Refs. \cite{Bocquet2010,Levine1975}.
The 'no-slip' conductance $G=I_{\rm no-slip}/E_0$ is the sum of the standard bulk term:
\begin{equation}
G_{\rm bulk}=2\mu e^2 C_{\infty} {\cal A}
\label{Gbulk}
\end{equation}

and a surface term \cite{Bikerman1933,Bocquet2010,Levine1975}, which can be written as:
\begin{equation}
G_{\rm surf,0}= 2 w\,\mu e \vert\Sigma\vert (1+\delta) \times {\chi \over {\sqrt{1+\chi^2}+1}}
\label{Gsurf0}
\end{equation}
where $\delta=1/(2\pi \ell_B\mu\eta)$ is a dimensionless parameter accounting for the electro-osmotic contribution to the conductance
($\eta$ the fluid viscosity); the parameter 
$\chi=1 / \kappa\ell_{GC}$ is proportional to the surface charge $\chi \propto \vert\Sigma\vert$.


We now consider the supplementary contribution induced by the slippage on the conductance, which is affected by the surface charge mobility. 
This can be written as:
\begin{equation}
I_{\rm slip} = {2w} \times \Sigma\, (v_--V_{slip})
\end{equation}

The slip velocity results from the electro-osmotic contribution to the flow, which we will consider  in details in section  III.D; its expression is given in Eq.(\ref{slip2}). 
Using the expression of the adsorbed charges velocity of Eq.(\ref{vion}), one deduces the slip-induced surface conductance as:
\begin{equation}
G_{\rm surf,1} = 2 w \left[  {b_{\rm eff}\over \eta} \times  (1-\alpha_s)^2\, \Sigma^2 + e \mu_w \lvert \Sigma \rvert \right]
\label{Gs1}
\end{equation}
where $\mu_w=1/(\lambda_s+\lambda_w)$ is the surface electro-phoretic mobility of the adsorbed ions. 

The first term is similar to expected slip contribution to the surface conductance, as for example discussed
in Ref. \cite{Bocquet2010}, but here corrected for the mobility of the adsorbed ions. The second term corresponds to the electrophoretic contribution of the mobile ions.

Altogether, the conductance is the sum of the three previous terms in Eqs.(\ref{Gbulk}), (\ref{Gsurf0}), (\ref{Gs1})
\begin{equation}
G=G_{\rm bulk}+G_{\rm surf,0}+G_{\rm surf,1}
\label{finalG}
\end{equation}

\subsection{Streaming currents}

We now consider the electric current induced by a flow in absence of any applied electric field. The flow is controlled by a pressure gradient $\nabla p$ applied on a channel. A Poiseuille flow builds up in and carries the charges in the system.

The electric current generated by the flow is then defined as:
\begin{equation}
I_{stream} = w\times \int_0^h e(C_+-C_-)(z)\,V(z) dz + 2w\times \Sigma v_-
\end{equation}
where the last term is the contribution of the mobile adsorbed layer.

Within the diffuse layer close to the surface, the flow can be approximated as $V(z)\simeq \dot\gamma_w (z+b_{\rm eff})$,
where the value for the shear rate at the surface, $ \dot\gamma_w$, is obtained in terms of the pressure drop
as $ \dot\gamma_w = {h\over 2\eta} (-\nabla p)$; $b_{\rm eff}$ is the effective slip length introduced above.
Similarly, the mobile ion velocity is defined in terms of the fluid slip velocity, $v_-=\alpha_s \times V_{slip}$,
with $\alpha_s={\lambda_s\over \lambda_s+\lambda_w}$. Accordingly we have 
$v_-=\alpha_s \times {h\over 2\eta} (-\nabla p)\times b_{\rm eff}$.
\\
\\
{
This allows to rewrite the streaming current as:
\begin{eqnarray}
I&_{stream} = &{{\cal A}\over \eta} (-\nabla p) \times \biggl[ \int_0^{h/2} {e\over 4\pi\ell_B}\left(-{d^2 \Phi \over dz^2}\right) (z+b_{\rm eff})dz \nonumber \\
&&+ \alpha_s\,\Sigma\,  b_{\rm eff}\biggr]
\end{eqnarray}
}


%
with ${\cal A}=wh$ the channel cross-sectional area.

Integrating the first term by parts { in the limit of small Debye layer, $h/\lambda_D \to \infty$} leads to:
\begin{equation}
I_{stream} = -{{\cal A}\over \eta} (-\nabla p) \times \biggl[ \epsilon\, \psi_0 +(1- \alpha_s)\,\Sigma\,  b_{\rm eff}\biggr]
\label{Istream}
\end{equation}

The corresponding streaming mobility takes the form:
\begin{equation}
\mu_{EO} = - {1\over \eta}\bigl[ \epsilon\, \psi_0 +(1- \alpha_s)\,\Sigma\,  b_{\rm eff}\bigr]
\label{muEO}
\end{equation}

The second term includes the mobile ion contribution, in addition to the slip induced contribution similar to Ref. \cite{Bocquet2010}.
As for the conductance, the  surface charge mobility is shown to correct the apparent surface charge  contribution to the streaming mobility by a factor $1-\alpha_s$.

\subsection{Electro-osmosis}

In the same geometry as the previous section, we now consider the fluid transport induced by a tangential electric field $E_0$.
Projected on the $x$-direction, the Stokes equation for a fluid element takes the form:
\begin{equation}
0=\eta \nabla^2 {V} - \mathbf{\nabla_x} p + e ( C_+-C_-) \times E_0
\end{equation}
which, using the Poisson's equation, can be rewritten as:
\begin{equation}
\eta {d^2 V \over dz^2} = -{e\over 4\pi\ell_B}\left(-{d^2 \Phi \over dz^2 }\right)\times E_0
\end{equation}

Integrated once ({ from $z$ to $h/2$ in the thin Debye layer limit}), one gets:
\begin{equation}
\eta {d V \over dz}(z) = -{e\over 4\pi\ell_B}\left(-{d \Phi \over dz }\right)\times E_0
\label{firstint}
\end{equation}
%

so that :
\begin{equation}
 {V_{EO}-V_{slip}} = -{\epsilon \psi_0\over \eta}\times E_0
\end{equation}
where we define the electro-osmotic velocity as { $V_{EO}=V(h/2)$, in the thin Debye layer limit}.

Crucially, the slip velocity results from the modified boundary condition in Eq.(\ref{slip}), 
taking into account the surface electric contribution to the slip boundary condition.
Combined with Eq.(\ref{firstint}), one obtains:
\begin{equation}
V_{slip}= 
-{ (1-\alpha_s)\Sigma b_{\rm eff}\over \eta}  E_0
\label{slip2}
\end{equation}

Gathering contributions one gets:
\begin{equation}
V_{EO}=-{1\over \eta} \left[{\epsilon \psi_0}+(1- \alpha_s)\,\Sigma\,  b_{\rm eff}\  \right]\times E_0 
\label{EO}
\end{equation}

Inspecting Eqs.(\ref{Istream}) and (\ref{EO}), one can verify that the Onsager symmetry, which imposes equality of the electro-osmotic and streaming mobility, is indeed satisfied, as it should.
This is an important validation for the modified hydrodynamic boundary condition, in particular of  the additional term accounting for the electric
field in the interfacial layer, which is central in the derivation of the electro-osmosis. To our knowledge, 
this modification of the hydrodynamic boundary condition is usually not taken into account.

\section{Diffusio-osmotic flow and currents}

In this section we explore specifically diffusio-osmotic transport, by which a flow is driven under a salt concentration gradient, as well as its consequence on the electric current induced by salinity gradients. Such interfacial transport phenomena play a central role in the context of blue energy harvesting \cite{Siria2017}.

\subsection{Diffusio-osmotic flow under salt gradient}

We first extend the calculation of the diffusio-osmotic flow in Ref. \cite{Prieve} to account for the combined effects of finite ion mobility and interfacial slippage. 
A salt concentration gradient $\nabla C_\infty$ is applied along the $x$-direction far from the surface. The Debye layer is much smaller than the distance over which $C_\infty(x)$ varies, so that one may consider ions to have relaxed locally towards equilibrium in the $z$-direction, 
$C_\pm(x,z)=C_\infty(x) \exp(\mp \Phi(z))$, 
with $\Phi$ the above solution of the PB equation.
The Stokes equation for a fluid element takes the form:
\begin{equation}
0=\eta \nabla^2 \mathbf{V} - \mathbf{\nabla} p + e ( C_+-C_-) (-\mathbf{\nabla} \psi) 
\end{equation}

Projecting the Stokes equation along the direction $z$ perpendicular to the surface,  one may integrate the pressure field to obtain
$p(x,z)=p_\infty + 2 k_BT\, C_\infty(x) \left[ \cosh \Phi(z) -1\right]$. 
Substituting this result in the $x$ component of the Stokes equation yields the equation for the velocity field:
\begin{equation}
\eta {d^2 V \over dz^2} = 2 k_BT\,  \left[ \cosh \Phi(z) -1\right] \times {d  C_\infty\over dx}
\label{Vx}
\end{equation}

This equation can be integrated twice and using the
effective slip boundary condition in Eq.(\ref{slip}) (here, with a vanishing electric field), one obtains the slip velocity of the fluid at the surface:
{
\begin{equation}
V_{slip}=- b_{\rm eff} {2 k_BT \over \eta} \,  \left[\int_0^{h/2} \left[ \cosh \Phi(z) -1\right] dz \right] \times {d  C_\infty\over dx}
\end{equation}
}
The integral can be computed exactly { in the thin Debye layer limit} using the change of variable $z(\Phi)$ which can be written, thanks to Eq.(\ref{PB}), as:
$dz = - {d\Phi / \kappa\sqrt{2  \left[ \cosh \Phi(z) -1\right] }}$.
This leads to:
\begin{equation}
V_{slip}=-  {2 b_{\rm eff} \lambda_D  \over \eta} \,  \left[ \cosh {\Phi_0\over 2} -1 \right] \times 2 k_BT {d  C_\infty\over dx}
\label{Vslip1}
\end{equation}
with $\Phi_0=e\psi_0/k_BT$ the dimensionless surface potential. 
Using $\cosh {\Phi_0\over 2} -1 = \sqrt{1+\chi^2}-1$, one
obtains the final expression of the slip velocity:

\begin{equation}
V_{slip}= 
{ k_B T \over 2\pi \eta\, \ell_B }
  {b_{\rm eff} \over \lambda_D } \,  \left[ \sqrt{1+\chi^2}-1\right]\times {-\nabla_x  \log C_\infty}
\label{Vslip}
\end{equation}
Using once more the Stokes equation, one can obtain the full diffusio-osmotic (plug) velocity profile far from the surface,
in the form
\begin{equation}
V_\infty = D_{DO} \times {-\nabla_x  \log C_\infty}
\end{equation}
where the diffusio-osmotic mobility (with the dimension of a diffusion coefficient) is now  the sum of the slip  and the bulk contributions:
\begin{eqnarray}
&D_{DO}& = { k_B T \over 2\pi \eta\, \ell_B }
 \Biggl( { b_{\rm eff} \over \lambda_D } \,  \left[ \sqrt{1+\chi^2}-1\right]  \nonumber \\
 &&+ \ln\left[ {\sqrt{1+\chi^2}+1\over 2}\right]\Biggr)
\end{eqnarray}
with $\chi$ the parameter previously defined as the ratio of the Debye to the Gouy-Chapman lengths $\chi(\Sigma)=1/\kappa \ell_{GC}$.

This expression of the diffusio-osmosis coefficient with mobile surface charges reveals that  the mobility of the adsorbed ions enters only via the modification of the effective slip length and -- in contrast to electro-osmosis -- there is no reduction of the `apparent' surface charge by the surface charge mobility. This can be rationalised by the fact the diffusio-osmosis arises from an equilibrium on the direction transverse to the one of the surface charges mobility.  Furthermore,
the contribution proportional to the slip length is basically the equivalent of a Marangoni flow induced by a surfactant gradient.

{Note that if the assumption of equal cation and anion mobilities is relaxed, then a supplementary ``chemiphoretic'' contribution to the diffusio-osmotic flow results from the so-called diffusional electric field which builds up under a salinity gradient and takes the form 
$$ E_{\rm diff}={k_BT\over e} \beta \times {(-\nabla_x  \log C_\infty})$$
with $\beta={\mu_+-\mu_-\over {\mu_++\mu_-}}$.
This will induce a supplementary electro-osmotic contribution to the diffusio-osmotic current. Using Eq.(\ref{EO}) -- which accounts for the modification of the hydrodynamic boundary condition under the supplementary electric field -- this leads to a supplementary
contribution to the diffusio-osmotic mobility in the form
\begin{equation}
D_{\rm DO}^{supp}=- \beta {k_BT\over e} \left[{\epsilon \psi_0\over \eta}+(1- \alpha_s)\,\Sigma\,  {b_{\rm eff}\over \eta}\  \right]
\end{equation}
}

\subsection{Generated osmotic current}

As for streaming currents, the diffusio-osmotic flow will convey ions in the electrical double layer and produce a corresponding electric current by convection of the ions in excess. 
This effect was first discussed by Fair and Osterle \cite{Fair1971} in the context of the reverse electro-dialysis
phenomenon and was measured experimentally in boron-nitride nanotubes \cite{Siria2013}. Here, the physisorbed ions will contribute to the induced diffusio-osmostic current by a supplementary contribution.

We still consider the rectangular channel described before, with dimensions larger than the Debye screening length (thin diffusive
layer limit). In this geometry the diffusio-osmotic current is defined as:
{
\begin{equation}
I_{DO}=2w \left[ \int_0^{h/2}dz\, e(C_+-C_-)\, V(z) + \Sigma \times v_-\right]
\label{IDOdef}
\end{equation}
where the last contribution accounts for the contribution of the mobile ions to the diffusio-osmotic current.
}
In this equation, the bulk fluid velocity $V$ follows from Eq.(\ref{Vx}), while the ion velocity in the 
interfacial layer is related to the slip velocity according to Eq.(\ref{vion}).
 { In the following we consider the limit of small Debye layer, $h/\lambda_D \to \infty$.
 }

Let us compute first the bulk contribution to the diffusio-osmotic current, {\it i.e.} the first term in Eq.(\ref{IDOdef}). Using Poisson's equation
$C_+-C_-=-\partial_z^2 \Phi/4\pi\ell_B$, this term can be rewritten as:
\begin{equation}
I_{DO}^{(1)}= 2w\, {e\over 4\pi\ell_B}\,  \int_0^{\infty} dz\, (-\partial_z^2 \Phi)\, V(z)
\end{equation}
This integral can then be integrated twice by parts to yield:
\begin{eqnarray}
&I_{DO}^{(1)}&= 2w\, {e\over 4\pi\ell_B}\, \biggl\{\partial_z \Phi(0)\, V(0) - \Phi(0) \,{dV\over dz}(0) \nonumber \\
&& - \int_0^\infty dz\, \Phi(z)\,  \left[ \cosh \Phi(z) -1\right] \times{2 k_BT\over\eta} \,{d  C_\infty\over dx} \biggr\}
\end{eqnarray}
where we used Eq.(\ref{Vx}).
The latter Stokes equation is also used to calculate the shear-rate at the surface
\begin{equation}
{dV\over dz}(z=0) = - {2k_BT\over \eta} \int_0^\infty dz\,  \left[ \cosh \Phi(z) -1\right] \times {d  C_\infty\over dx} 
\end{equation}
Gathering expressions, one obtains:
\begin{eqnarray}
&I_{DO}^{(1)}&= 2w\, e \Sigma_n V_{slip}   \nonumber \\
&&+2w\, {e}\,{k_BT\over 2\pi\eta\,\ell_B} \,{d  C_\infty\over dx}  
 \int_0^\infty dz\,  \left(  \Phi_0-\Phi(z) \right) \left[ \cosh \Phi(z) -1\right] \nonumber \\
\end{eqnarray}
with the slip velocity given by Eq.(\ref{Vslip}). 
The integral can be calculated using the same change of variables as above to obtain the  result:
%
%

\begin{equation}
I_{DO}^{(1)}=  2w\, e \Sigma_n V_{slip} + 2w\, {e}\, {k_BT\over \pi\eta\,\ell_B}   {1\over \kappa}\, \left(2\sinh {\Phi_0\over 2} -\Phi_0\right) {\nabla_x  C_\infty} 
\end{equation}

Rewriting this equation in terms of the surface charge (remembering that $\Sigma=-e\Sigma_n<0$) and using the relationship between the surface potential and the surface charge of Eq. (\ref{Surfpot}), one obtains:
\begin{eqnarray}
&I_{DO}^{(1)}&=  2w\times  (-\Sigma) { k_B T \over 2\pi \eta\, \ell_B } \biggl\{
  {b_{\rm eff} \over \lambda_D } \,  \left[ \sqrt{1+\chi^2}-1\right] \nonumber \\
&&  + \left(1 - { \sinh^{-1} \chi \over \chi} \right) \biggr\}
\times {-\nabla_x  \log C_\infty} \nonumber \\
\end{eqnarray}
with $\chi=1/\kappa\ell_{GC}$.


Now the second (surface) contribution to the diffusio-osmotic current in Eq.(\ref{IDOdef}) can be rewritten 
in terms of the fluid slip velocity:
\begin{equation}
I_{DO}^{(2)}=2w\,  \Sigma \times  \alpha_s \times V_{slip}
\end{equation}
with $\alpha_s=\lambda_s / (\lambda_s+\lambda_w)$ defined in terms of the physiosorbed ion friction.
Using the expression for the slip velocity obtained above in Eq.(\ref{Vslip}), 
one gets:
\begin{eqnarray}
&I_{DO}^{(2)}&=2w\,  \Sigma   \times
 {2 b_{\rm eff} \alpha_s \lambda_D  \over \eta} \,  \left[ \sqrt{1+\chi^2}-1\right] \nonumber \\
 &&\times 2 k_BT (-\nabla_x \log C_{\infty})
\end{eqnarray}

Overall the corresponding mobility $K_{\rm osm}$, defined as:
$$I_{DO} = K_{\rm osm} \times (-\nabla_x \log C_{\infty}),$$
thus takes the
form $K_{osm} = K^{(1)}_{\rm osm}+ K^{(2)}_{\rm osm}$ with:
\begin{equation}
K^{(1)}_{\rm osm}=  {2w} \, \times (-\Sigma)  {k_BT\over 2\pi\eta\,\ell_B}\left(1 - { \sinh^{-1} \chi \over \chi}\right) 
\label{K1}
\end{equation}
and
\begin{equation}
K^{(2)}_{\rm osm}= {2w }\times (-\Sigma) { k_B T \over 2\pi \eta\, \ell_B } (1-\alpha_s)
  {b_{\rm eff} \over \lambda_D } \,  \left[ \sqrt{1+\chi^2}-1\right] 
\end{equation}
with $\alpha_s= {\lambda_s\over \lambda_s+\lambda_w}$. 
We remind that the effective slip length is a decreasing function of the surface charge, as $b_{\rm eff} (\Sigma)=b_0/(1+\beta_s \Sigma_n)$.

For high surface charge, the second  contribution $K^{(2)}_{\rm osm}$ reduces to the simple form:
\begin{equation}
K^{(2)}_{\rm osm} = {2w\over L}  {e\Sigma_n^2}   \times
 { b_{\rm eff} (1-\alpha_s)} {k_BT \over \eta } 
 \end{equation}
 Since $b_{\rm eff}\sim 1/\Sigma_n$, this term scales as $K^{(2)}_{\rm osm} \propto \Sigma_n$, like the first contribution $K^{(1)}_{\rm osm}$.

{Finally, if the assumption of equal cation and anion mobilities is again relaxed, a supplementary contribution to the diffusio-osmotic current
arises from the induced diffusional electric field as discussed above. This leads to a supplementary contribution to the mobility $K_{\rm osm}$ which can be written as:
\begin{equation}
K^{\rm supp}_{\rm osm}= G \times {k_BT\over e}\times \beta,
\end{equation}
where the expression for the conductance $G$ is given in Eq.(\ref{finalG}). In this expression the ion mobility is taken as the average $\mu={1\over 2}(\mu_++\mu_-)$.
}

\section{Discussion}

\subsection{Small versus large surface charge}


The various transport coefficients depend strongly on the surface charge with a change of behaviour characterized by the parameter $\chi = \lambda_D / \ell_{GC}$, with a transition occurring for $\chi =1$. We can rewrite this condition in terms of surface charge density ($\Sigma=-e \Sigma_n$) and salt concentration $C_{\infty}$: $\chi={(2\pi \Sigma_n \ell_B) / \sqrt{8\pi \ell_B C_{\infty}}}$, so that the limit between the two regimes is fixed by the threshold 
${\Sigma_n^2\ell_B/ C_{\infty} } \approx 1$.
This suggests the introduction of a dimensionless charge number: $${\mathbb C}_e={\Sigma_n^2\ell_B\over C_{\infty} }$$
Interestingly, {Eq.(\ref{Surfpot}) (which correspond to the Grahame equation \cite{grahame1947electrical}) shows that} ${\mathbb C}_e>1$ corresponds to the onset of non-linearities in the Poisson-Boltzmann framework, so that the large charge/low concentration behaviour of the transport coefficients are not properly predicted by a linearised Debye-H\"uckel theory.
For a given surface charge, the concentration threshold below which surface charge effect will show up is therefore
$[C_{\infty}]_{\rm threshold}=\Sigma_n^2\ell_B$. For $\Sigma=0.1$ C/m$^2$, $[C_{\infty}]_{\rm threshold} \approx 0.4$ M, so that the transport coefficient are strongly affected by non-linear electrostatic effects as soon as the salt concentration is below 0.4 M.
On the other hand, for $\Sigma=0.01$ C/m$^2$, $[C_{\infty}]_{\rm threshold} \approx 4$ mM so that non-linear electrostatic effect only apply for very low concentrations, and the low charge regime apply for most of the concentrations window. 

The consequences on transport are considerable:

\noindent$\bullet$ For ${\mathbb C}_e >1$ ({\it i.e.} $\Sigma_n>\sqrt{C_{\infty}\over \ell_B} $), which corresponds to a large surface charge or low salt concentrations regime, strong non-linear effects occur: the surface contributions to transport conductance and diffusio-osmotic current scale linearly in the surface charge $\sim \Sigma^1$, while the electro-osmotic mobilities scale as a sum of a constant and a surface charge logarithmic term $\mu \sim {\rm cst} + {\rm log} \ \vert \Sigma\vert$.\\
\noindent$\bullet$ For ${\mathbb C}_e <1$ ({\it i.e.} $\Sigma_n<\sqrt{C_{\infty}\over \ell_B} $), which corresponds to the small surface charge  or high salt concentration regime, the surface contributions to transport coefficients scale as $\Sigma^n$ with $n=1$ (electro-osmotic mobility), $n=2$ (conductance and diffusio-osmotic mobilities) and $n=3$ (diffusio-osmotic currents).



Accordingly the value of the surface charge has a considerable effect on the transport phenomena, in particular in the small salt concentration regime (quantified by ${\mathbb C}_e <1$). The linear scaling in surface charge for electrokinetic phenomena highlighted for large charges does change to quadratic or even cubic dependences in surface charge for small charges: such contributions are accordingly vanishingly small.

For example in the case of ionic current induced under salinity gradients, the mobility
$K_{\rm osm}$ scales as $K_{\rm osm}\sim \Sigma^3$ from Eq.(\ref{K1}) at low concentration, while $K_{\rm osm}\sim \Sigma$ for large charge. Accordingly the diffusio-osmotic current is expected to be extremely small for small charge but strongly enhanced for large charges. 
A linear scaling of $K_{\rm osm}$ in surface charge was measured experimentally in boron-nitride nanotubes in Ref. \cite{Siria2013}, for which very high surface charge was measured. The different behaviours of the transport coefficient as a function of the surface charge are displayed in the Table I. for high and low surface charges and with and without mobile charges.

As a last comment, we mention that the discussion of large versus small charge limiting regime becomes particularly subtle when the surface charge takes its origin in charge regulation mechanisms. Indeed it was shown recently \cite{Secchi2016PRL,Biesheuvel2016,Manghi2018,Uematsu2018}
that in this case,  the surface charge is expected to scale with the salt concentration via a scaling dependence, in the form
$\Sigma\propto C_{\infty}^{p}$, with an exponent $p\approx 1/3-1/2$ depending on the salinity regime.
In particular for $p=1/2$,  ${\mathbb C}_e$ becomes independent of the concentration, so that the electrolyte  remains either in the strongly non-linear regime, or in the weakly non-linear regime, whatever the value of salinity $C_{\infty}$. 
This affects directly the expectations for the amplitude of the various transport coefficients, whose mobility $\mu$ will scale as a power law as a function of salinity like $\mu\propto (C_{\infty})^{np}$ (depending on the regime and transport phenomena under investigation).

\begin{table*}[t]
\begin{tabular}{|c|c|c|c|c|}
  \hline
   & \multicolumn{2}{c|}{Low surface  charge} & \multicolumn{2}{c|}{High surface charge}  \\
     \cline{2-5}
     & & & & 
     \\[-0.9em]
   & without mobile charges & with mobile charges & without mobile charges & with mobile charges  \\
     \hline 
     & & & & 
     \\[-0.9em]
$\mu_{\rm EO}$ & $\Sigma + b_{\rm 0}\Sigma$ & $\Sigma + (1-\alpha_s)b_{\rm eff}\Sigma$ &$cst+\log\lvert \Sigma \rvert + b_{\rm 0}\Sigma$ &$cst+\log\lvert \Sigma \rvert+(1-\alpha_{\rm s})b_{\rm eff}\Sigma$ \\
     \hline 
     & & & & 
     \\[-0.9em]
$D_{\rm DO}$ & $\Sigma^2 + b_{\rm 0}\Sigma^2$ & $\Sigma + b_{\rm eff}\Sigma^2$ &$cst+\log\lvert \Sigma \rvert + b_{\rm 0}\Sigma$ &$cst+\log\lvert \Sigma \rvert+b_{\rm eff}\Sigma$ \\  
     \hline 
     & & & & 
     \\[-0.9em]
$K_{\rm DO}$ & $\Sigma^3 + b_{\rm 0}\Sigma^3$ & $\Sigma^3 + (1-\alpha_s)b_{\rm eff}\Sigma^3$ &$\Sigma+b_{\rm 0}\Sigma^2$ &$\Sigma + (1-\alpha_{\rm s})b_{\rm eff}\Sigma^2$ \\
     \hline 
     & & & & 
     \\[-0.9em]
$G_{\rm surf,0}$ & $\Sigma^2 $& $\Sigma^2 $ &$\Sigma $ &$\Sigma$ \\
     \hline 
     & & & & 
     \\[-0.9em]
$G_{\rm surf,1}$ & $b_{\rm 0}\Sigma^2 $& $\mu_w \lvert \Sigma \rvert + (1-\alpha_s)^2b_{\rm eff}\Sigma^2$ &$b_{\rm 0}\Sigma^2 $ &$\mu_w \lvert \Sigma \rvert + (1-\alpha_s)^2b_{\rm eff}\Sigma^2$ \\
     \hline 
\end{tabular}
\caption{Table showing the main scalings of the different dependences on surface charge and slippage of the transport coefficient, with and without mobile surface charges, for both the low and high surface charge regimes. We remind that the effective slip length $b_{\rm eff}$ depends on the surface charge according to Eq.(\ref{slip}). 
\\
}
\end{table*}


\subsection{Mobile ion contribution to the transport coefficient}

The derivation of the transport coefficient in the case of mobile surface charge exhibits in most cases a moderating mechanism which tends to reduce the surface effects, with the noticeable exception of diffusio-osmosis and for the conductance which we discuss in the next section. In particular, it appears through an effective reduced surface charge $(1-\alpha_s)\Sigma$, and via the reduction of the effective slip-length (since at high surface charge $b_0 \gg b_{\rm eff} \sim 1/\Sigma$). We remind that $1-\alpha_s=\lambda_w/(\lambda_s+\lambda_w)$ the ratio between the friction coefficient of the adsorbed ions on the surface to the total friction coefficient.

This can be seen in particular for streaming currents.  A large slip length would be expected to  induce a high value for the zeta potential, here defined as $\zeta = \psi_0 + (1-\alpha_s)\Sigma b_{\rm eff} / \epsilon$ . However, the mobility of the adsorbed ions will reduce the slippage effect via the $(1 - \alpha_s)$ coefficient.  In the limiting case of a perfectly mobile surface charge  $\alpha_s = 1$, the contribution due to slippage fully vanishes. Physically, (and in the absence of surface potential $\psi_0$) this situation would corresponds to a flow conveying a globally neutral electrolyte (hence no induced streaming current), separated into a negatively charged adsorbed layer and a positively charged Debye layer both translated at the same velocity. The same reasoning apply to the contribution of slippage for electro-osmosis and diffusio-osmotic currents.

It is accordingly crucial to obtain an estimate of the ion friction on the wall, in addition to the slip length, in order to rationalize the effect of slippage on the corresponding transport coefficients and disentangle the various contributions.

%
%

\subsection{Surface conductance versus Zeta potential: some hints to a puzzle ?}

It is worth noting that the moderating effect discussed above does not apply for all transport coefficients. Indeed the conductance has an additional positive contribution due to the ability of surface charges to move when an electric field is applied. 
As reported in Eq. (\ref{Gs1}), this term is {\it linear}  in surface charge: $G_{\rm s,e} = 2we\mu_w\lvert\Sigma\rvert$, with $\mu_w=1/(\lambda_w+\lambda_s)$ the adsorbed ion mobility.

{ In contrast, the electro-phoretic mobility of adsorbed ions appears in the streaming/electro-osmotic mobility through a coefficient $\lambda_w b_{\rm eff} \mu_w$.} This contribution involves a  prefactor which is different by nature since it is directly related to the slippage of both water ($b_{\rm eff}$) and that of the adsorbed charges (via $\lambda_w$).
This points that the link between surface conductance and charge may be not obvious and straightforward in some systems. 

This is particularly evident for small surface charges (${\mathbb C}_e <1$). Indeed
as $\Sigma\rightarrow 0$, the electrophoretic contribution to the surface conductance $G_{\rm s,e}$ remains linear
in $\Sigma$, while the standard surface conductance, {\it e.g.} $G_{\rm surf,1}$ and $G_{\rm surf,0}$, are quadratic
in $\Sigma$ for low charge. 
{In this limit, the surface conductance is merely dominated by the electrophoretic contribution of the adsorbed ions:}
\begin{equation}
G_{\rm s,e} = 2we\mu_w\lvert\Sigma\rvert
\end{equation}
while the electro-osmotic mobility is then: 
\begin{equation}
\mu_{\rm EO} \approx {\lvert \Sigma \rvert \lambda_D \over \eta} \left(1 + {\lambda_w \over \lambda_w+\lambda_s}\times {b_{\rm eff} \over \lambda_D}\right)
\end{equation}

which for large slippage (say, for carbon interface) can be expressed as a function of the dominant electro-phoretic surface conductance: 
\begin{equation}
\mu_{\rm EO} \approx {G_{\rm s,e} \over 2 w e} \times { \lambda_w b_{\rm eff} \over \eta }
\end{equation}
This expression makes clearly visible that the two quantities are not obviously related, because they are coupled via the product of the slip length to the ion friction on the interface. Mobility of charges should accordingly be a supplementary effect to include in order to rationalize  the origin of some discrepancies between zeta potential and conductance, which was exhaustively discussed in several works \cite{Lyklema1994,Bonthuis2016}.
 Note that this equation also provides a direct way to quantify $\lambda_w$, the ion-wall friction coefficient, by measuring in this regime both conductance and electro-osmotic mobility. 
%
%

{
\subsection{Combined effects of fixed and adsorbed charges}
In this  section we briefly extend the previous analysis by considering the combined effects  of coexisting mobile and fixed charges at the surface on transport coefficients. We denote  as $\Sigma_t$ the total surface charge. It is the sum of the mobile $\Sigma_m$ and fixed $\Sigma_0$ surface charges so that $\Sigma_t = \Sigma_m  + \Sigma_0$. The parameter $\chi$ is now proportional to the total surface charge $\Sigma_t$, and we now give the different expressions of the transport coefficient.
The derivations follows similar steps as in the previous calculations and we give here only the main results.

\noindent{\it Modified hydrodynamic boundary condition--}
The modified hydrodynamic boundary condition involves the mobile contribution to the surface charge, as~:
\begin{equation}
 b_{\rm eff} \times \partial_z V\biggl\vert_{wall} = V_{slip}  - \alpha_s{b_{\rm eff}\over \eta} \times {\Sigma_m}\,(-\partial_x\psi)\biggl\vert_{wall} 
\label{slip_fix}
\end{equation}

where the effective slip length is now:
\begin{equation}
b_{\rm eff}= {b_0 \over 1 + \beta_s \times \rvert \Sigma_m /e \lvert}
\label{bslip_fix}
\end{equation}

\vskip0.2cm
\noindent{\it Conductance --}
The total conductance $G$ is the sum of a bulk term, which remain unchanged, and two surface terms that are now given by:
%

\begin{equation}
G_{\rm surf,0}= 2 w\,\mu e \vert\Sigma_t\vert (1+\delta) \times {\chi \over {\sqrt{1+\chi^2}+1}}
\label{Gsurf0_fix}
\end{equation}

\begin{equation}
G_{\rm surf,1} = 2 w \left[  {b_{\rm eff}\over \eta} \times  [\Sigma_0 + \Sigma_m (1-\alpha_s)]^2 + e \mu_w \lvert \Sigma_m \rvert \right]
\label{Gs1_fix}
\end{equation}
where $\mu_w=1/(\lambda_s+\lambda_w)$ is the surface electro-phoretic mobility of the adsorbed ions. 

\vskip0.2cm
\noindent{\it Streaming or electro-osmotic mobility --}
The electro-osmotic mobility takes now the form:
\begin{equation}
\mu_{EO} = - {1\over \eta}\bigl[ \epsilon\, \psi_0 +(\Sigma_0 + \Sigma_m (1-\alpha_s))\,  b_{\rm eff}\bigr]
\label{muEO_fix}
\end{equation}
\vskip0.2cm

\noindent{\it Diffusio-osmosis --}
The diffusio-osmotic mobility is not modified except via the dependency of $b_{\rm eff}$ with the mobile surface charge density $\Sigma_m$ and $\chi$ which depends on the total surface charge $\Sigma_t$.
In contrast, the diffusio-osmotic current mobility is explicitly modified and now takes the form:
$K_{osm} = K^{(1)}_{\rm osm}+ K^{(2)}_{\rm osm}$ with:
\begin{equation}
K^{(1)}_{\rm osm}=  {2w} \, \times \vert \Sigma_t\vert  {k_BT\over 2\pi\eta\,\ell_B}\left(1 - { \sinh^{-1} \chi \over \chi}\right) 
\label{K1_modif}
\end{equation}
and
\begin{equation}
K^{(2)}_{\rm osm}= {2w }\times \vert \Sigma_0 + \Sigma_m (1-\alpha_s)\vert { k_B T \over 2\pi \eta\, \ell_B }
  {b_{\rm eff} \over \lambda_D } \,  \left[ \sqrt{1+\chi^2}-1\right] 
\end{equation}
\vskip0.5cm
Altogether, it is interesting to note that the transport coefficients gather differential contributions originating in both the fixed and  mobile part of the surface charge, with a complex interplay between the two. This also originates from the differential dependence of transport on the electrostatic and dynamic contributions of the mobile charges. This may actually allow to separate the two contributions from the study of all transport coefficients for a given interface.
}
\subsection{Consequences on experimental measurements}

Carbon nanotubes represent a very interesting system to explore the consequences of the surface charge mobility on the electrokinetics.
These systems were shown to highlight very large, radius dependent, slippage \cite{Secchi2016Nature}, while exhibiting a substantial surface charge
\cite{Secchi2016PRL}, as inferred from conductance measurements.
The latter was conjectured to originate in the adsorption of hydroxide ions at the carbon surface. Recent {\it ab initio} simulations actually confirmed the physisorption of OH$^-$ on carbon surfaces, while strong covalent bonding on boron-nitride surfaces \cite{Grosjean}. Furthermore the hydroxide ions were shown to keep a large lateral surface mobility on carbon surfaces, even when physisorbed.
From the above expressions for the various mobilities, the slip induced contributions to the transport mobilities are expected to be dominant for carbon nanotubes, because the slip length was measured to be very large, up to hundreds of nanometers (increasing with decreasing diameter) \cite{Secchi2016Nature}. 
However, the dependence of the slip length with salinity remains up to now unexplored, in particular in the context of the prediction for the charge dependence of the slip length derived here, Eq.(\ref{bslip}). Furthermore, as emphasized previously, the amplification effect due to interfacial slippage may be counteracted by the finite mobility of the physisorbed ions (via the $1-\alpha_s$ term in the various mobilities). The two effects, slippage and interfacial ion mobility, are intertwined and it is not {\it a priori} obvious to predict the amplitude of the corresponding slippage term in the transport mobility.
This suggests to explore the various transport coefficients in carbon nanotubes and investigate their variations with salt concentration, pH, and tube radius, in order to disentangle the various contributions to the transport mobilities. Only the comparison of all transport coefficients would allow to extract unambiguously the various coefficients at play ({\it e.g.} slip lengths, surface mobility, friction coefficients, etc.).





Another phenomenon of interest is the so-called blue energy harvesting, whereby electric current is generated by salinity gradients, see \cite{Siria2017} for a review. 
As highligthed in Ref.~\cite{Huang2016}, hydrodynamic slippage can strongly enhance the salinity-gradient power conversion. However, in this work we show that the slip-length cannot be considered without the effect of adsorbed mobile charges that tends to reduce the harvesting efficiency. For example, carbon nanotubes and their large slip length could be at first sight a good candidate for energy conversion despite their relatively low surface charge. However, their adsorbed charge mobility (that remains to be measured) could reduce the harvesting efficiency. Again, only a detailed investigation of the various transport phenomena would allow to confirm the contribution of the various phenomena at play in the electric current generation under salinity gradients.



%
%

\begin{acknowledgements}
\noindent LB acknowledges funding from the European Union's H2020 Framework Programme/ERC Advanced Grant agreement number 785911 - {\it Shadoks}. The authors acknowledge support from ANR {\it Neptune}.
We thank M.-L. Bocquet, B. Grosjean, R. Vuilleumier and A. Siria for highlighting discussions.
\end{acknowledgements}


\begin{thebibliography}{99}

\bibitem{Hunter} R. J. Hunter {\it Foundations of Colloid Science} (Oxford Univ. Press, New York, 1991).

\bibitem{Bocquet2010} L. Bocquet, E. Charlaix {\it Chem. Soc. Rev.}  \textbf{39}, 1073-1095. (2010)

\bibitem{Siria2017} A. Siria, ML. Bocquet, L. Bocquet, {\it Nature Reviews Chemistry}, \textbf{1}, 0091. (2017).

\bibitem{Joly2004} L. Joly, C. Ybert, E. Trizac, L. Bocquet
 {\it Phys. Rev. Lett.} {\bf 93} 257805 (2004).

\bibitem{Joly2006} L. Joly, C. Ybert, E. Trizac, L. Bocquet
 {\it J. Chem. Phys.} {\bf 125} 204716 (2006).
%

\bibitem{Balme2015}
S. Balme, F. Picaud, M. Manghi, J. Palmeri, M. Bechelany, S. Cabello-Aguilar, A. Abou-Chaaya, P. Miele, E. Balanzat, J.M. Janot, 
{\it Sci. Rep.} {\bf 5}, 10135 (2015).

\bibitem{Huang2016} D. Rankin and D.M. Huang, {\it Langmuir} {\bf 32} 3420 (2016).
\bibitem{Huang2008} D.M. Huang, C. Cottin-Bizonne, C. Ybert, L. Bocquet, {\it Langmuir} {\bf 24} 1442 (2008)

\bibitem{Bonthuis2012} D.J. Bonthuis, R.R. Netz,  {\it Langmuir} {\bf 28} 16049 (2012).

\bibitem{Bonthuis2016} D.J. Bonthuis, Y. Uematsu, R.R. Netz,  {\it Phil. Trans. R. Soc. A} {\bf 374} 20150033 (2016).


\bibitem{Secchi2016PRL}  E. Secchi, A. Nigu\`es, L. Jubin, A. Siria \& L. Bocquet, {\it Physical review letters}, 116(15), 154501. (2016)

\bibitem{Biesheuvel2016} P. M. Biesheuvel, M.Z. Bazant, 
{\it Phys. Rev. E} {\bf 94}, 050601 (2016)


\bibitem{Uematsu2018} Y. Uematsu, R.R. Netz, L. Bocquet, D. Bonthuis, {\it J. Phys. Chem. B} {\bf 122}, 2992 (2018). 

\bibitem{vanroij2018} B.L. Werkhoven, J.C. Everts, S. Samin, R. van Roij, {\it Phys. Rev. Lett.} {\bf 120}, 264502 (2018).


\bibitem{Fleck2002} C. Fleck, R.R. Netz, H.H. von Gr{\"u}nberg, {\it Biophysical journal}, {\bf 82} 76 (2002)


\bibitem{Andelmann2018} Y. Avni, D. Andelmann, R. Podgornik, {\it Current Opinion in Electrochemistry}, {\bf 13} 70-77 (2018)

\bibitem{Joly2014} L. Joly, F. Detcheverry, A.-L. Biance, {\it Phys. Rev. Lett.} {\bf 113} 088301 (2014).

\bibitem{Lyklema1994}
J. Lyklema, {\it Colloids and Surfaces A}, {\bf 92} 42 (1994).

\bibitem{Secchi2016Nature} E. Secchi, S. Marbach, A. Nigu\`es, D. Stein, A. Siria \& L. Bocquet. {\it Nature}, 537(7619), 210. (2016)

\bibitem{Grosjean} B. Grosjean, M.-L. Bocquet, R. Vuilleumier, submitted (2018)

\bibitem{Mouterde2018} T. Mouterde, {\it et al.}, accepted for publication in {\it Nature} (2018).

\bibitem{Andelman} D. Andelman,  {\it Handbook of biological physics} (Vol. 1, pp. 603-642). (North-Holland, 1995).

\bibitem{Levine1975} S. Levine, J. R. Marriott and K. Robinson, J. Chem. Soc.,
Faraday Trans. 2, 711 (1975).

\bibitem{Bikerman1933}
J.J. Bikerman, {\it Z. Phys. Chem. A}, {\bf 163} 378 (1933)


\bibitem{Prieve} D.C. Prieve, J.L. Anderson, J.P. Ebel, M.E. Lowell, {\it J. Fluid Mech.} {\bf 148} 247-269 (1984)

\bibitem{Fair1971} J. C. Fair and J. F. Osterle, {\it J. Chem. Phys.}, {\bf 54}, 3307 (1971).

\bibitem{Siria2013} A. Siria, P. Poncharal, A.-L. Biance, R. Fulcrand, X. Blase, S. Purcell, L. Bocquet,
{\it Nature} {\bf 494}, 455-458 (2013).

\bibitem{grahame1947electrical} D. C. Grahame, {\it Chemical reviews} {\bf 3} 441 (1947)
%

\bibitem{Manghi2018} M. Manghi, J. Palmeri, Y. Khadija, F. Henn, V. Jourdain, {\it Phys. Rev. E} {\bf 98} 012605 (2018).



\end{thebibliography}
\end{document}